\documentclass[10pt]{article}
\usepackage{graphicx}

\def\Title#1{\begin{center} {\Large #1 } \end{center}}
\def\Author#1{\begin{center}{ \sc #1} \end{center}}
\def\Address#1{\begin{center}{ \it #1} \end{center}}

\newcommand\pubblock{\rightline{\begin{tabular}{l} Proceedings of the Fifth Annual LHCP\\ \pubnumber\\
         \pubdate  \end{tabular}}}

\newenvironment{Abstract}{\begin{quotation} \begin{center}
            \large ABSTRACT \end{center}\bigskip
      \begin{large}}{\end{large} \end{quotation}}

\newenvironment{Presented}{\begin{quotation} \begin{center} 
             PRESENTED AT\end{center}\bigskip 
      \begin{center}\begin{large}}{\end{large}\end{center} \end{quotation}}

\def\beq{\begin{equation}}
\def\eeq#1{\label{#1}\end{equation}}
\def\eeqn{\end{equation}}

\def\beqa{\begin{eqnarray}}
\def\eeqa#1{\label{#1}\end{eqnarray}}
\def\eeqan{\end{eqnarray}}

\let\bar=\overbar

\def\Dslash{\not{\hbox{\kern-4pt $D$}}}
\def\dslash{\not{\hbox{\kern-2pt $\del$}}}

\def\msb{{\bar{\ssstyle M \kern -1pt S}}}

\usepackage{color}

\newcommand\pubnumber{ ATL-PHYS-PROC-2017-103 }

\newcommand\pubdate{\today}

\def\affiliation{
On behalf of the ATLAS Experiment, \\
Department of Physics \\
Calabria University, Arcavacata di Rende, 87036, Cosenza, Italy}

\begin{document}

\large
\begin{titlepage}
\pubblock

\vfill
\Title{  Production of vector bosons in association with jets in ATLAS  }
\vfill

\Author{ Evelin Meoni  }
\Address{\affiliation}
\vfill

\begin{Abstract}
Measurements of the production of jets in association with a W/Z boson in proton–proton collisions are presented using data collected by the ATLAS experiment at LHC at $\sqrt s$ = 8 and 13 TeV. Several kinematic regimes are explored  with various approaches to probe different aspects of these processes.

The differential cross sections of a Z boson in association with jets with $p_T>$ 30 GeV and $|y|<$2.5 at $\sqrt s$ = 13 TeV  are measured in a fiducial phase space, probing strong interactions  that completely dominate in these processes, 
while measurements of a W boson in association with at least two jets 
at high $p_T$ and high di-jet invariant mass, where the electroweak production is enhanced, 
are performed with $\sqrt s$ = 8 TeV data. 
Angular distributions in W+jets events with high $p_T$ jets  
 are also measured at $\sqrt s$ = 8 TeV focusing on small angular separation between the jets and the W decay products, where contributions from real W emission are expected large. 
Finally a measurement of the splitting scales occurring in the $k_t$ jet-clustering algorithm is presented for final states containing a Z boson at  $\sqrt s$ = 8 TeV. This measurement based on charged-particle track information constitutes a complementary approach to study jet properties.   


Data distributions of all the measurements are corrected for detector effects and compared with state-of-the-art predictions.  

\end{Abstract}
\vfill

\begin{Presented}
The Fifth Annual Conference\\
 on Large Hadron Collider Physics \\
Shanghai Jiao Tong University, Shanghai, China\\ 
May 15-20, 2017
\end{Presented}
\vfill
\end{titlepage}
\def\thefootnote{\fnsymbol{footnote}}
\setcounter{footnote}{0}
%

\normalsize 


\section{Introduction}

The measurement of the cross sections for the production of hadronic jets in association with a W/Z boson at LHC provides a stringent test of perturbative quantum chromodynamics (pQCD) and can be used to improve understandings on the parton distribution function (PDF). Moreover, since these processes form frequent backgrounds for searches of new physics, their detailed measurement is a mandatory step in the discovery program at LHC.

Although the production of a W/Z boson in association with two or more jets is dominated by processes involving strong interactions, the large cross section for W-boson production allows to explore the region of  high transverse momentum ($p_T$) jets and high di-jet mass ($m_{jj}$), where  the electroweak (EW) production is enhanced.
This topology constitutes an important background for Higgs studies in vector boson fusion production.

At high energies, also real emission of weak bosons in di-jet events can contribute significantly to inclusive W+jets processes.  This kind of contributions is expected to be large  at small angular separations between
a high $p_T$ jet and the lepton coming from the W decay. 
A good understanding of the real W emission process is important not only for W+jets measurements at high $p_T$ , but also for vector-boson scattering measurements and  for  multijets measurements at high di-jet mass. In addition, this process has a high potential to mimic the highly Lorentz-boosted top quark production, which makes it important also for New Physics searches.

Finally while properties of the jets can be studied directly using the jet momenta in W/Z+jets events, a complementary approach can be adopted by studying the jet production rates at different resolution scales via measurement of the splitting scales occurring in the $k_t$ jet-clustering algorithm.
 The measurement is based on charged-particle track information, which is measured with excellent precision in the $p_T$ region relevant for the transition between the perturbative and the non-perturbative regimes. 

This contribution presents a review of the latest measurements of jet production in events with a W/Z boson, where aforesaid aspects are explored using data collected by the ATLAS experiment \cite{atlas} at LHC at $\sqrt s$ = 8 and 13 TeV. 

\begin{figure}[htb]
\centering
\includegraphics[height=3.3in]{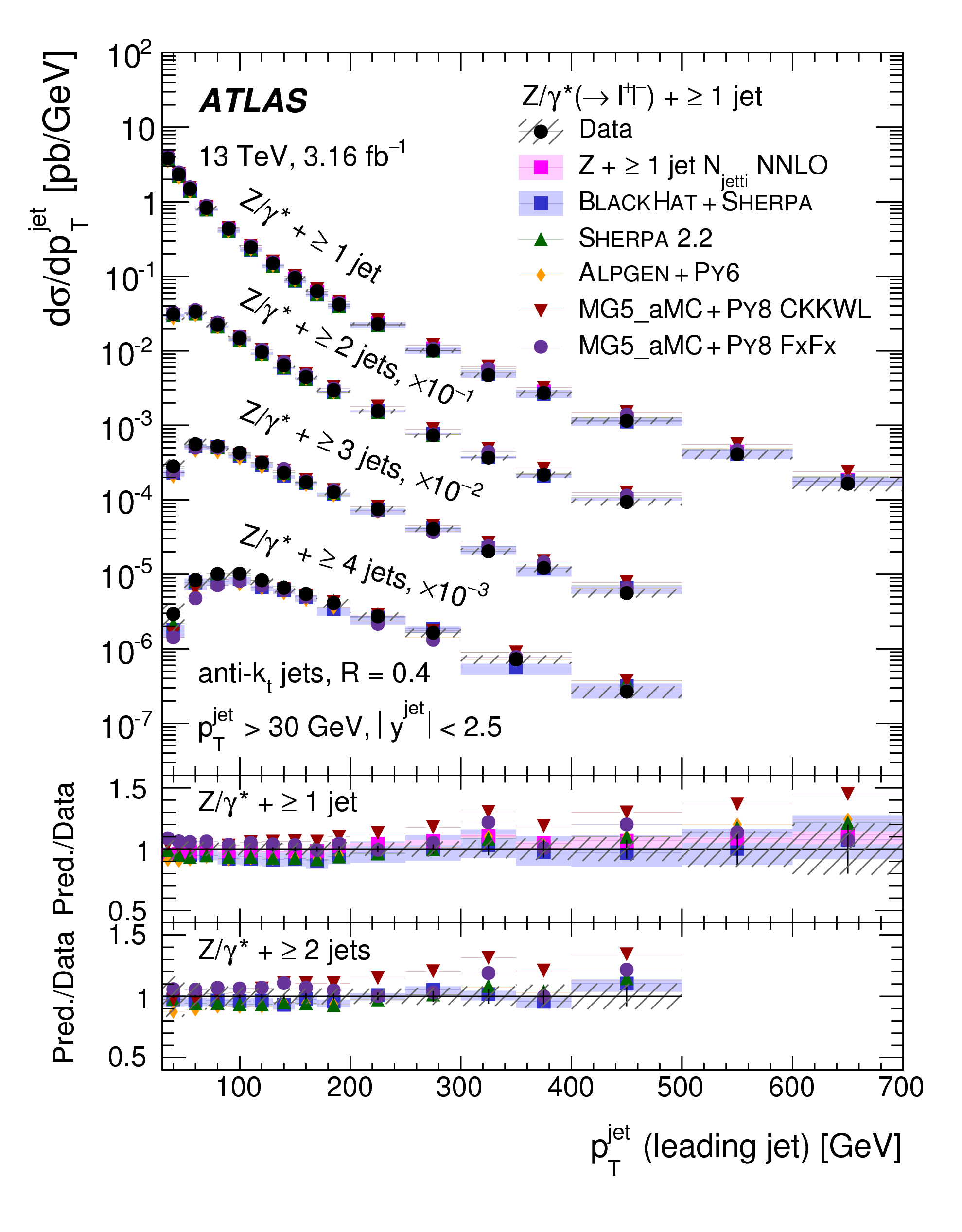}
\includegraphics[height=3.3in]{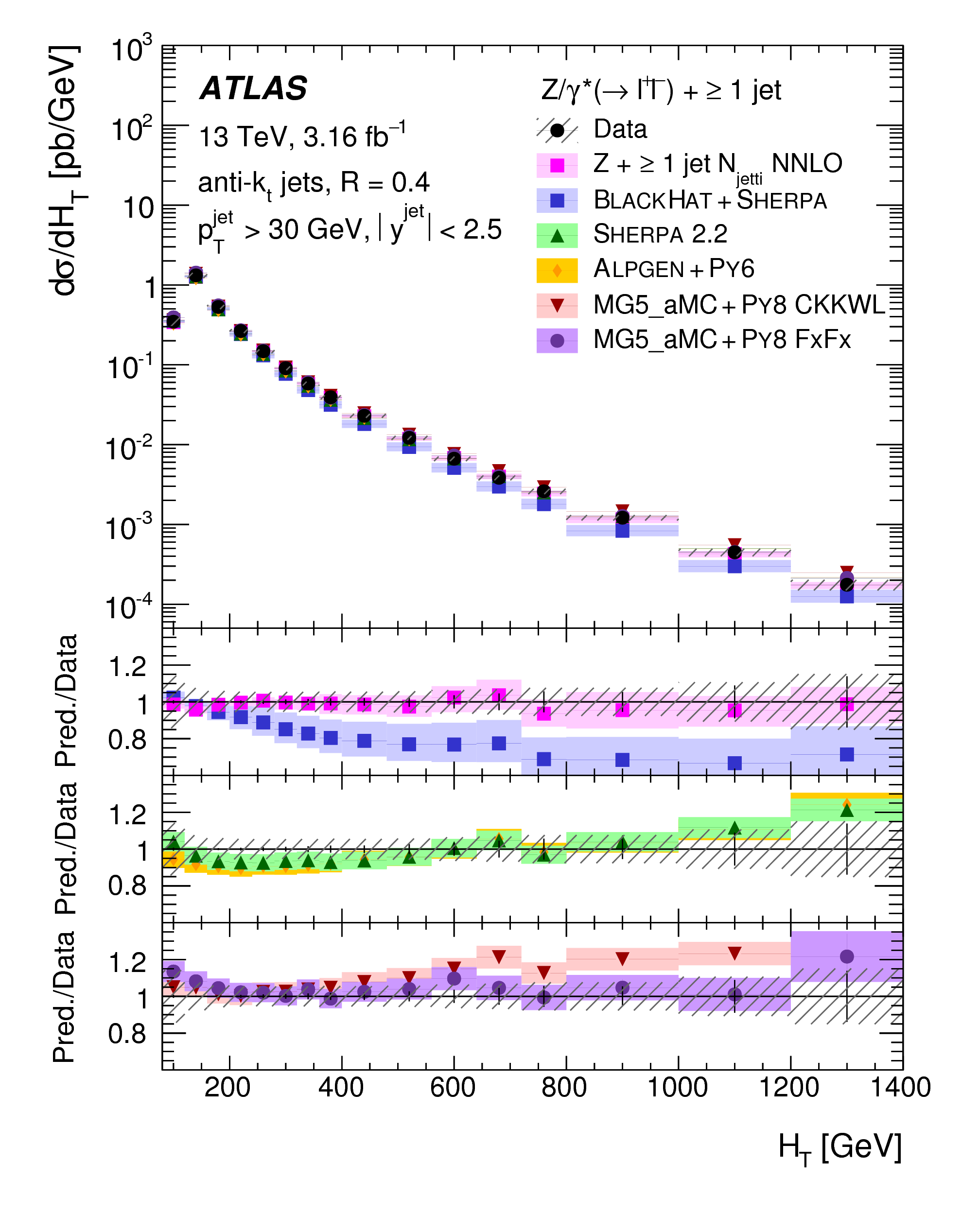}
\caption{Measured cross section as a function of the leading jet $p_T$ for inclusive Z+$\ge$ 1,2,3,4 jet events (left)
and as a function of $H_T$ for inclusive Z+$\ge$ 1 jet events (right). The data are compared to the predictions from Z+$\ge$ 1 jet N$_{jetti}$ NNLO, BLACKHAT+SHERPA, SHERPA 2.2, ALPGEN+PY6, MG5$\_$aMC+PY8 CKKWL, and MG5$\_$aMC+PY8 FxFx. The error bars correspond to the statistical uncertainty, and the hatched bands to the data statistical and systematic uncertainties (including luminosity) added in quadrature \cite{zjets}.}
\label{fig:figure1}
\end{figure}
\begin{figure}[htb]
\centering
\includegraphics[height=3in]{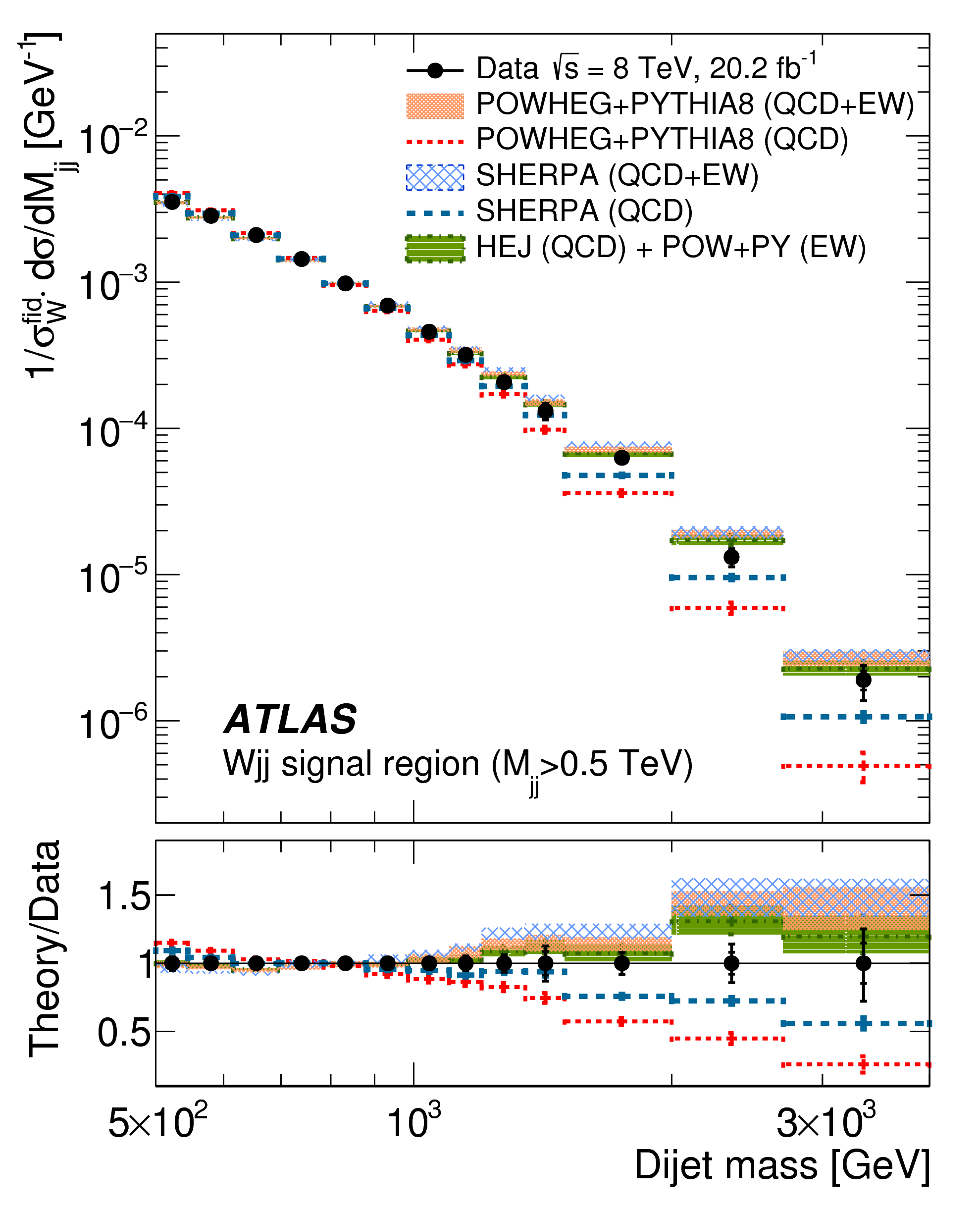}
\includegraphics[height=3in]{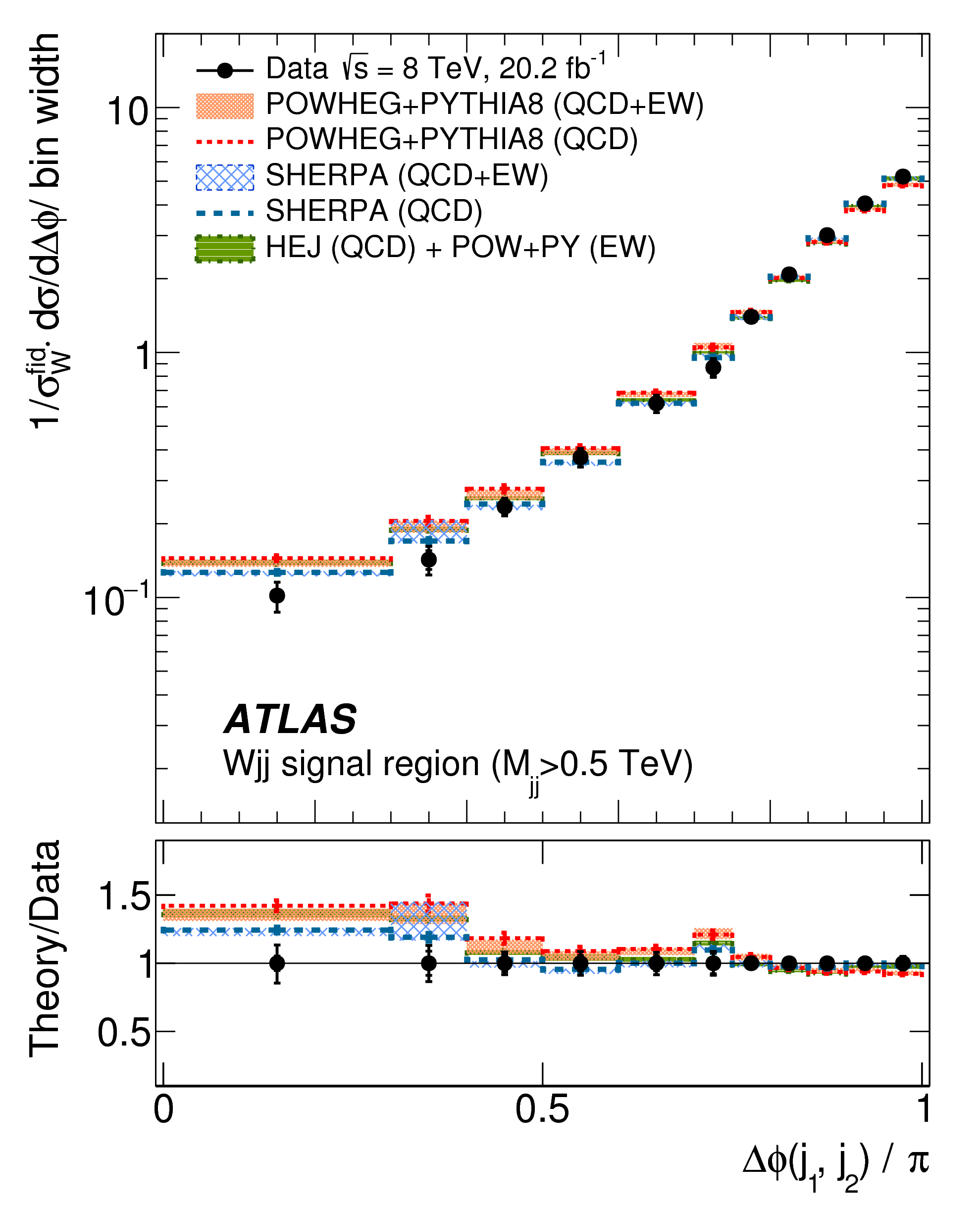}
\caption{Unfolded normalized differential Wjj production cross sections as a function of $m_{jj}$ (left) and as a function of $\Delta \phi_{jj}$ (right) in the signal region defined as described in the text. Both statistical (inner bar) and total (outer bar) measurement uncertainties are shown, as well as ratios of the theoretical predictions to the data (the bottom panel in each distribution) \cite{ewwjets}.}
\label{fig:figure2}
\end{figure}
\begin{figure}[htb]
\centering
\includegraphics[height=3in]{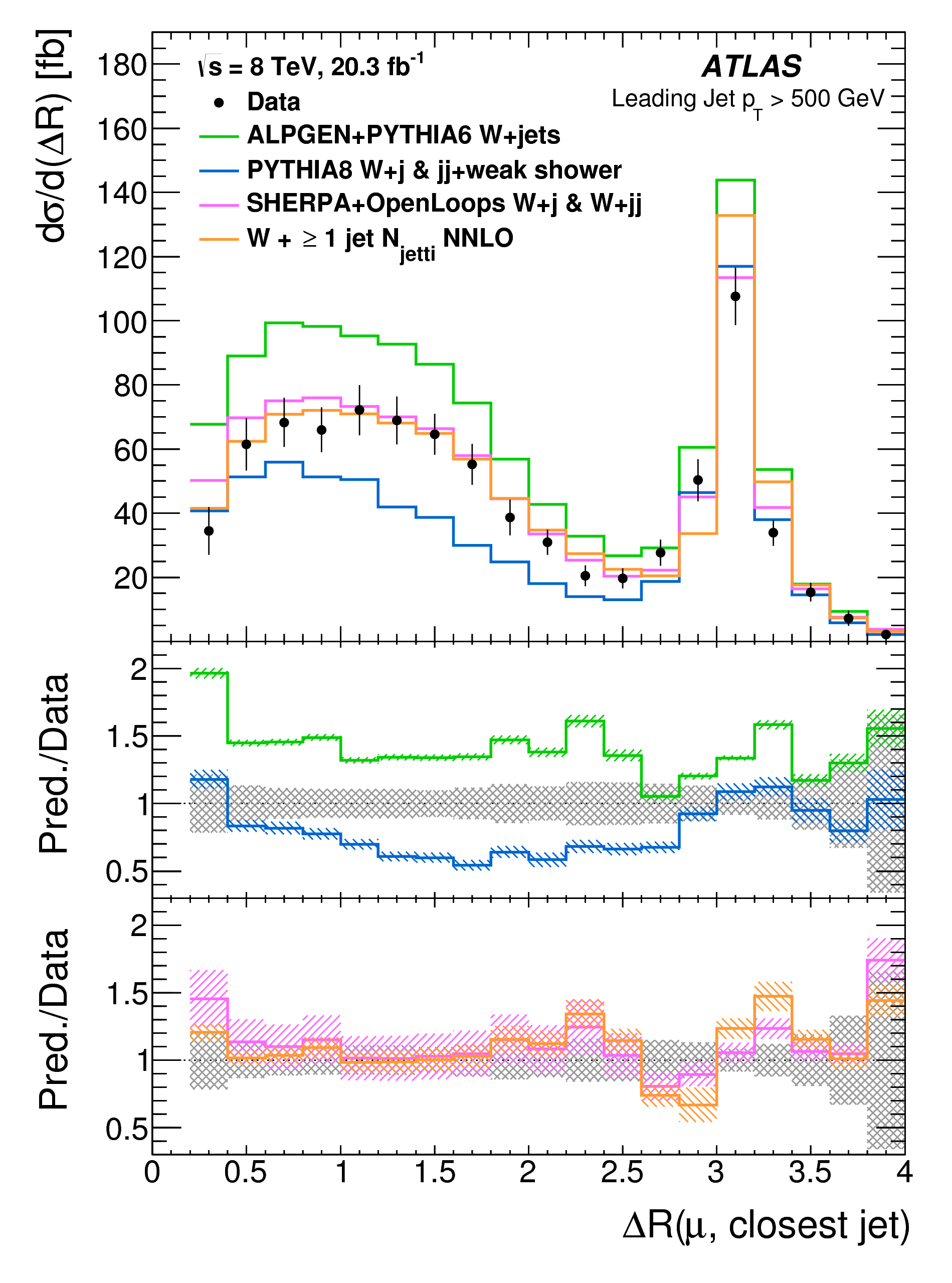}
\caption{ Measured W+jets cross section as a function of the angular separation between the muon and the closest jet along with several predictions from theory calculations. The lower panels show the ratio of the theory predictions to the unfolded data. The error bars in the upper panel and the grey shaded error bands in the lower ratio panels are the sum of the statistical and systematic uncertainties in the measurement. The shaded error band on the ALPGEN+PYTHIA6 calculation is statistical uncertainty, the band on the PYTHIA8 calculation is statistical and PDF uncertainties and those on the SHERPA+OpenLoops and the W + ≥ 1 jet 
N$_{jetti}$  NNLO calculations are scale uncertainties \cite{wjetsang}. }
\label{fig:figure3}
\end{figure}
\begin{figure}[htb]
\centering
\includegraphics[height=2in]{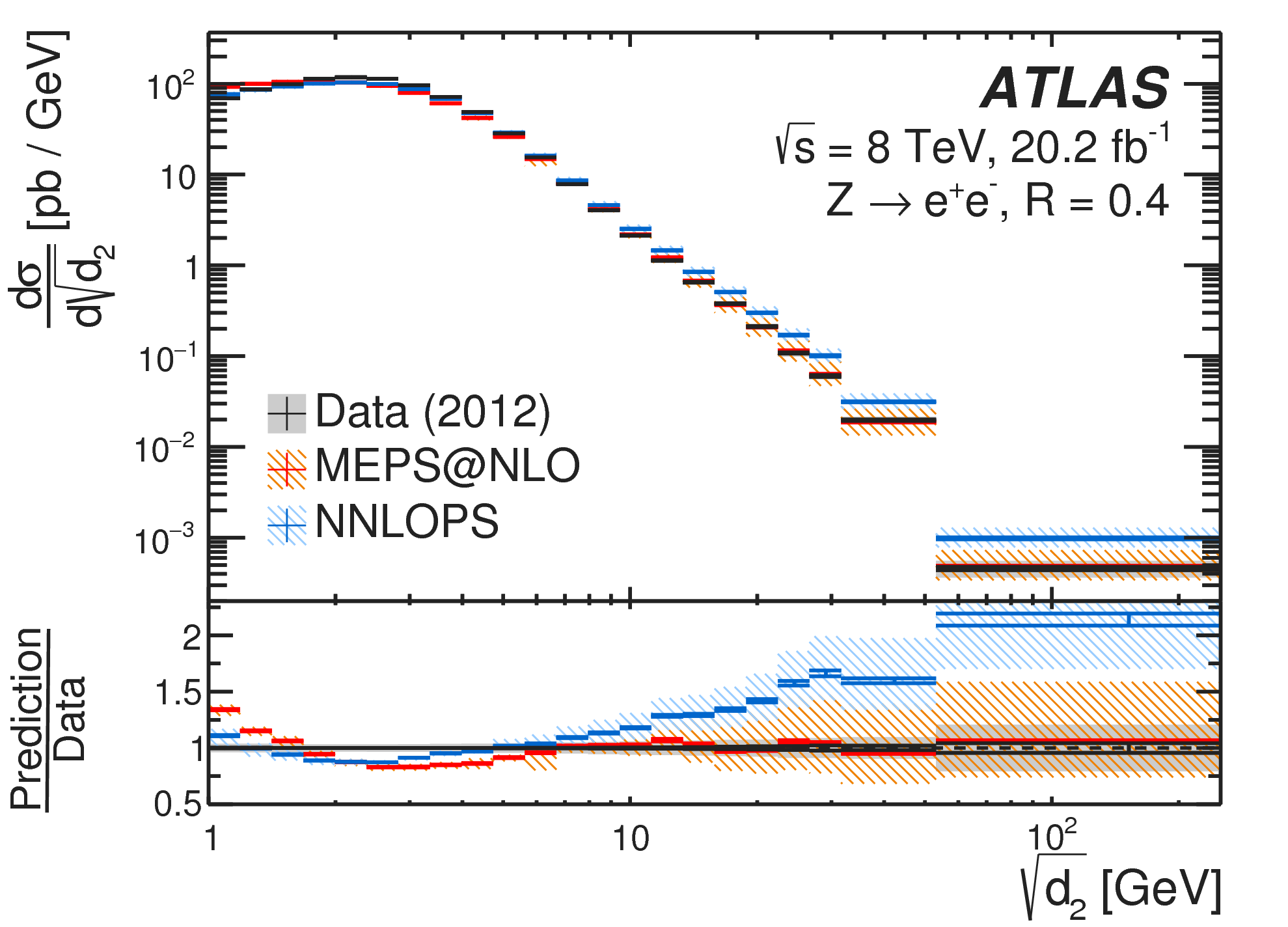}
\includegraphics[height=2in]{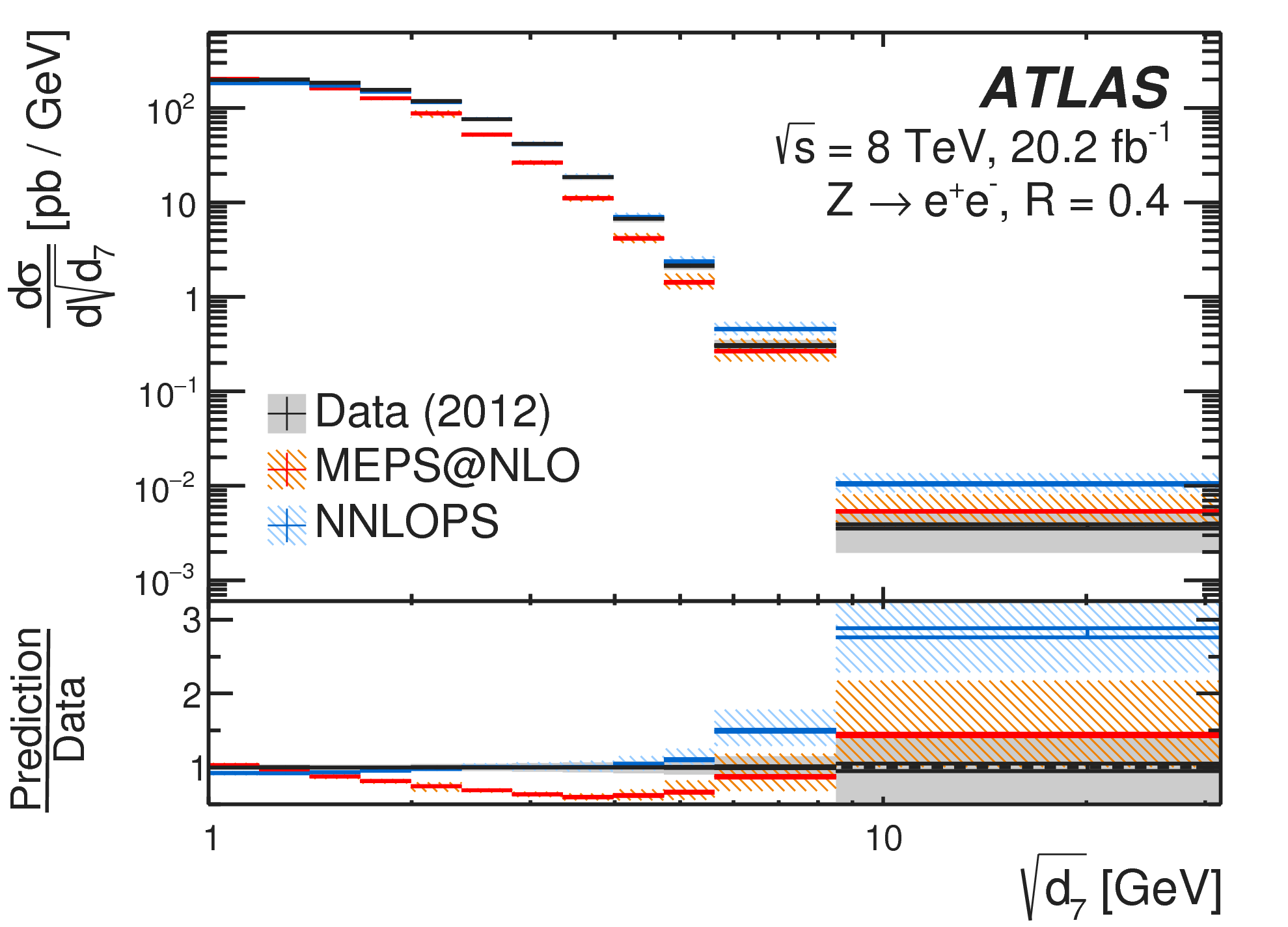}
\caption{ Unfolded distribution for  splitting scale $\sqrt d_2$ (left) and $\sqrt d_7$ (right) in the electron channel using the jet-radius parameter R=0.4. The size of the error bars reflects the statistical uncertainty, while the combined statistical and systematic uncertainty is indicated by the grey band. Theoretical predictions from SHERPA with NLO multijet merging ("MEPS@NLO") and from POWHEG+PYTHIA 8 with NNLO matching ("NNLOPS") are displayed including error bands for the generator uncertainties \cite{zjetskt}.}
\label{fig:figure4}
\end{figure}

\section{Z+jets measurements at 13 TeV}

The cross-section measurement of the Z boson production in association with jets is performed  using 3.16 $fb^{- 1}$ of data collected by the ATLAS detector at $\sqrt s$ = 13 TeV \cite{zjets}.
The Z boson is identified using its decays to electron
or muon pairs ($Z \to \ell^+ \ell^-$ with $\ell=e$ or $\mu$). Jets are clustered using the anti-$k_t$ algorithm \cite{jets} with radius parameter R = 0.4.  The fiducial acceptance
region is defined by the following requirements: $p^l_T >$ 25 GeV, $|\eta^l| <$ 2.5, 71$ <m_{ll}<$111GeV, $p^{j}_T>$ 30 GeV, $|y^{j}| <$ 2.5 and $|\Delta R(l,j)|>$0.5. Differential cross sections are measured as a function of different variables.
 
Figure \ref{fig:figure1} (left) shows the differential cross sections for the combination of electron and muon channels as a function of the leading jet $p_T$ for Z+$\ge$ 1,2,3,4 jet events. The Z+$\ge$1 jet N$_{jetti}$ NNLO prediction models the spectrum for the Z+$\ge$1 jet events well. NLO calculation from BlackHat+SHERPA, SHERPA 2.2 Monte Carlo (MC) and MG5$\_$aMC+Py8 FxFx MC, which are based on NLO matrix elements, and LO Alpgen+PY6 MC predictions agree with data within systematic uncertainties. The LO MG5$\_$aMC$+$PY8 CKKWL MC models a too-hard jet $p_T$ spectrum for $p_T^j>$ 200 GeV, this can be interpreted as the indication that the dynamic factorization and renormalization scales used in the generation are not appropriate for the full jet $p_T$ range.

Figure \ref{fig:figure1} (right) shows the differential cross sections as a function of the scalar sum $H_T$ of the transverse momenta of  leptons and jets for Z+$\ge$ 1 jet events. The predictions from SHERPA 2.2, Alpgen+PY6 and MG5$\_$aMC+PY8 FxFx describe well the $H_T$ distribution. The prediction from MG5$\_$aMC+PY8 CKKWL describes well the turn-over in the softer part of the $H_T$ spectrum, but overestimates the contribution at large values of $H_T$, in line with the overestimate of the cross sections for hard jets. The fixed-order Z+$\ge$ 1 jet prediction from BlackHat+SHERPA underestimates the cross section for values of $H_T >$  300 GeV due to the missing contributions from events with higher parton multiplicities, which for large values of $H_T$ constitute a substantial portion of the data. Agreement is recovered by adding higher orders in perturbative QCD, as demonstrated by the good description of $H_T$ by Z+$\ge$ 1 jet 
N$_{jetti}$ NNLO.

\section{W+2 jets measurements in high mass regime at 8 TeV}

The cross-section measurement of the W boson production in association with at least 2 energetic jets at high
 di-jet mass is performed  using 20.2 $fb^{- 1}$  of data collected by the ATLAS detector at $\sqrt s$ = 8 TeV \cite{ewwjets}.
The W  boson is identified via leptonic decay ($W \to \ell \nu$ with $\ell=e$ or $\mu$). The fiducial acceptance
region is defined by the following requirements: $p^l_T >$25 GeV, $|\eta^l| <$2.5,  $p^{\nu}_{T}>$20 GeV, W's transverse mass $m_T>$40 GeV, $p^{j1}_T>$80 GeV and $p^{j2}_T>$60 GeV both  with $|y^{j}|<$4.4, $m_{jj}>$500 GeV and $|\Delta R(l,j)|>$0.3. Two additional requirements are applied to further enhance the electroweak production against the strong one: exactly 1 lepton and 0 jets in the central region defined by the directions of  the two leading jets. 

Figure \ref{fig:figure2} (left)  shows the normalised differential cross sections as a function of $m_{jj}$. LO SHERPA MC and NLO POWHEG+PYTHIA MC give a satisfactory description of data when both strong  and electroweak  processes are included in the simulation.

Figure \ref{fig:figure2} (right) shows the normalised differential cross sections as a function of the angular azimuthal separation between the two leading jets ($\Delta \phi_{jj}$). $\Delta \phi_{jj}$ is not sensitive to EW and QCD separation, but it is a discriminant variable in Higgs measurements, moreover it probes the interplay between matrix element (radiation at large angles) and parton shower (soft collinear radiation) implemented in MCs. Good agreement between the data and all predictions is seen, with a slight tendency for predictions to overestimate the relative rate at small angles.

\section{Angular measurements in W+jets at 8 TeV}
The differential cross section of the production of a W boson in association with at least one high-$p_T$ jet is measured using 20.3 $fb^{- 1}$  data collected at $\sqrt s$ = 8 TeV \cite{wjetsang}. The measurement is performed as a function of the distance between the muon from the W decay ($W \to \mu \nu$) and the closest jet, $\Delta R(\mu, j)$, requiring \\ $p^\mu_T>$ 25 GeV, $|\eta^\mu|<$ 2.4. In order to enrich the collinear production of W+jets the leading jet is required to have $p_T >$ 500 GeV.
Figure \ref{fig:figure3} shows the 
result. The comparison of the data to ALPGEN+PYTHIA6 MC  shows good shape agreement to within uncertainties, except at 
very low $\Delta R$, but ALPGEN+PYTHIA6 MC predicts a significantly higher integrated cross section. The comparison to PYTHIA8 MC at high $\Delta{R}$, where it is dominated by back-to-back W+jets production in which the W boson is balanced by the hadronic recoil system, shows much better agreement. At smaller $\Delta R$, where the collinear process dominates, neither the shape nor the overall cross section agree. The comparisons to SHERPA+OpenLoops and W+$\ge$ 1 jet N$_{jetti}$ NNLO show much better agreement across the entire distribution.

\section{$k_T$ splitting measurement in Z+jets at 8 TeV}
The differential cross sections of the Z+jets production are measured using 20.2 $fb^{- 1}$  data collected by the ATLAS detector at $\sqrt s$ = 8 TeV as functions of the splitting scales occurring in the $k_T$ clustering algorithm using charged-particle tracks as inputs \cite{zjetskt}.
The splitting scales of jets are constructed using an infrared-safe clustering algorithm based on a sequential combination of the input momenta, which approximates QCD evolution. The splitting scale $d_k$ is defined for a given iteration of the algorithm at which a number of the input momenta drops from k + 1 to k, i.e. the zeroth-order splitting scale ($d_0$)
corresponds to the $p_T$ of the leading $k_T$-jet.
 This measurement is sensitive to the hard perturbative modeling at high scales and to soft hadronic activity at lower scales. 

Figure \ref{fig:figure4} shows the differential cross section as a function of the low-order splitting scales, $d_2$, and of the higher-order splitting scales, $d_7$. Neither of the generators, namely SHERPA (‘MEPS$@$NLO’) and DY$@$NNLO+POWHEG+PYTHIA 8 (‘NNLOPS’), provides a fully satisfactory description of the experimental data.
For lower-slitting scales both predictions overshoot the data significantly at low values and underestimate data in the peak region , while at high values  the MEPS$@$NLO prediction agrees well with the data while the NNLOPS prediction systematically overestimates the cross section. The level of agreement of the NNLOPS predictions in the soft region is improved significantly for the higher-order splitting scales.

\clearpage
\newpage
\section{Conclusions}

The ATLAS experiment at LHC has a wide range of recent measurements of a vector boson with associated jets at $\sqrt s$ = 8 and 13 TeV.
These  measurements  are an important tool  for testing perturbative QCD. Moreover they constitute precious inputs 
to improve the modelling of MCs in different kinematical regions important for many measurements and for searches.
In addition to a Z+jets cross-section measurement in a typical phase space, also electroweak enhanced regions and
collinear regions are investigated in W+jets events.  Splitting scales in the $k_T$ algorithm are also measured
in Z+jets events as complementary approach to study jet production rates at different resolution scales in addition to direct studies of the jet properties.
Comparisons to a variety of MC generators and theoretical calculations show varying levels of agreement, highlighting the success of recent theoretical advances and the opportunity for further tuning.

\end{document}